\documentclass[a4paper]{article}
\usepackage{epsfig,amssymb,amsmath}

\def\beq{\begin{eqnarray}}
\def\eeq{\end{eqnarray}}
\def\bsp{\begin{split}}
\def\esp{\end{split}}
\def\lra{\longrightarrow}

\newcommand{\mb}[1]{{\mathbb #1}}
\newcommand{\mc}[1]{{\cal #1}}

\begin{document}

\title{\textbf{The Weyl tensor in Spatially Homogeneous Cosmological Models}}
\author{\textbf{John D. Barrow}\thanks{%
J.D.Barrow@damtp.cam.ac.uk}~ and\ \textbf{Sigbj\o rn Hervik}\thanks{%
S.Hervik@damtp.cam.ac.uk} \\
\\
DAMTP,\\
Centre for Mathematical Sciences,\\
Cambridge University, \\
Wilberforce Rd.,\\
Cambridge CB3 0WA, UK}
\date{\today}
\maketitle

\begin{abstract}
We study the evolution of the Weyl curvature invariant in all spatially
homogeneous universe models containing a non-tilted $\gamma $-law perfect
fluid. We investigate all the Bianchi and Thurston type universe models and
calculate the asymptotic evolution of Weyl curvature invariant for generic
solutions to the Einstein field equations. The influence of compact topology
on Bianchi types with hyperbolic space sections is also considered. Special
emphasis is placed on the late-time behaviour where several interesting
properties of the Weyl curvature invariant occur. The late-time behaviour is
classified into five distinctive categories. It is found that for a large
class of models, the generic late-time behaviour the Weyl curvature
invariant is to dominate the Ricci invariant at late times. This behaviour
occurs in universe models which have future attractors that are plane-wave
spacetimes, for which all scalar curvature invariants vanish. The
overall behaviour of the Weyl curvature invariant is discussed in
relation to the proposal that some function of the Weyl tensor or its
invariants should play the role of a gravitational 'entropy' for
cosmological evolution. In particular, it is found that for all
ever-expanding models the measure of gravitational entropy proposed by Gr{\o}n and Hervik increases at late times.
\end{abstract}

\section{Introduction}

What is the entropy of a gravitational field? Following the insights of
Hawking and Bekenstein \cite{bekenstein,hawking}, an answer has been only
been provided for a subset of situations characterised by static and
stationary spacetimes with event horizons. In particular, black holes are
found to be black bodies. They possess a well-defined entropy. They obey the
laws of equilibrium thermodynamics and, when perturbed, their fluctuations
obey the laws of close-to-equilibrium thermodynamics \cite{sciama}. Yet,
despite these developments it remains an unsolved problem whether an
analogous gravitational entropy can be defined for non-stationary spacetimes
 of the sort that characterise most relativistic
cosmologies that are considered as models for the past, present, and future
of our visible universe. Unfortunately, a solution of this problem appears
to involve the solution of a number of separate hard problems in
gravitation physics. It requires a rigorous measure of the probability of
different cosmological initial conditions to be found. It also requires us
to be sure that we have accounted for all the possible contributions to the
gravitational entropy: there may be quantum, classical, geometrical,
topological, thermal, and dynamical contributions that need to be quantified
and added \cite{Hu}. The nature of the states of zero and maximum entropy also need
to be defined and the influence of possible events like inflation must be
fully accounted for entropically, and all must be suitably coordinate
covariant. It has also been argued \cite{davies74,davies83} that
the whole concept of the arrow of time and thermodynamics relies upon the
concept of a gravitational entropy, and hence leads all the way back to
fundamental questions about the origin of time and the nature of the initial
state of the universe.

Gravitation has a tendency to amplify small inhomogeneities in the
distribution of matter. Yet, our present observations of the spectrum and
temperature isotropy of the microwave background radiation show that the radiation was very close to thermal equilibrium when it was
last--scattered, and that fluctuations from the Friedmann-Robertson-Walker
(FRW) spacetime metric are only of order $10^{-5}$. If there does exist a
gravitational entropy that grows with time to reflect the action of
gravitational instability then it should have the property that it vanishes
(or at least is small) for a homogeneous and isotropic metric and be maximal for a black
hole of the same total mass. This type of observation has led some
investigators to propose that any measure of cosmological gravitational
entropy should be proportional to the deviation of the universe from the FRW
model \cite{bt}. Hence, the present gravitational entropy of the universe is in some
sense small and so must have been even smaller in the past, perhaps
providing some rationale for assuming the initial state of the universe to
have zero entropy and hence be FRW up to minimal quantum gravitational
fluctuations.

Based on this observation, Penrose \cite{penrose} suggested that the Weyl
curvature invariant, given by 
\begin{equation}
C^{\alpha \beta \gamma \delta }C_{\alpha \beta \gamma \delta }=R^{\alpha
\beta \gamma \delta }R_{\alpha \beta \gamma \delta }-2R^{\alpha \beta
}R_{\alpha \beta }+\frac{1}{3}R^{2},  \label{weyl}
\end{equation}
should be very small near the initial singularity, and then
grow thereafter behaving as a proper measure of the gravitational entropy. This in all its editions, is what we will call the \emph{Weyl Curvature
Conjecture} (WCC). Wainwright and Anderson \cite{wa} had a different version
of the WCC. They expressed the conjecture in terms of the relative magnitude
of the Weyl and Ricci curvature invariants: 
\begin{equation}
P^{2}=\frac{C^{\alpha \beta \gamma \delta }C_{\alpha \beta \gamma \delta }}{%
R^{\mu \nu }R_{\mu \nu }}.  \label{weyl1}
\end{equation}
However, based on observations from two different universe models and taking
into account that the entropy should scale with the volume, Gr\o n and
Hervik \cite{wcc1}, \cite{wcc2} suggested that a better measure of a
gravitational entropy would be 
\begin{equation}
\mathcal{S}=\sqrt{|g|}P.  \label{S}
\end{equation}
Most other studies of the WCC \cite{ra,rothman,gw,Bon85,Husain,pl,GCW} use
the square of the Weyl scalar, $C^{\alpha \beta \gamma \delta }C_{\alpha
\beta \gamma \delta },$ or $P,$ as the putative gravitational entropy for
cosmological models. However, these candidates do not scale as the
volume and so
they do not capture correctly the dependence of a Weyl entropy on the volume.

We introduce the expansion-normalised shear: 
\[
\Sigma =\frac{3}{2}\frac{\sigma _{\mu \nu }\sigma ^{\mu \nu }}{\theta ^{2}}. 
\]
The shear tensor is a symmetric and trace-free tensor defined by 
\[
\sigma _{\mu \nu }=\theta _{\mu \nu }-\frac{1}{3}\theta (g_{\mu \nu }+u_{\mu
}u_{\nu }). 
\]
Here, $u_{\mu }$ is the normalised four-velocity of the comoving fluid; $%
\theta _{\mu \nu }\equiv u_{\mu ;\nu }$ is the expansion tensor of the
comoving fluid and $\theta =\theta _{~\mu }^{\mu }$ the volume expansion
rate. The shear tensor tells us how anisotropic the universe is; in
particular, $\Sigma =0$ if and only if the universe is isotropic. From the generalised
Friedmann equation ($8\pi G=c=1$) 
\beq
\frac 13\theta^2=\frac 12\sigma^{\mu\nu}\sigma_{\mu\nu}-\frac
12{}^{(3)}R+\rho+\Lambda. 
\label{friedmann}\eeq
Since the density, $\rho$, is positive, $\Sigma $ will be bounded
by 
\begin{equation}
0\leq \Sigma \leq 1  \label{sig}
\end{equation}
whenever the spatial Ricci-scalar is non-positive: ${}^{(3)}R\leq
0$. This inequality will be fulfilled in
all of the homogeneous models,  except in
the type IX and the Kantowski-Sachs closed universes.

One of the aims of this paper will be to investigate the quantity $%
\mathcal{S}$ for all homogeneous models, and determine whether or not it has the
appropriate behaviour for a gravitational entropy. Hence, we will try to
determine whether or not $\mathcal{S}$ increases during the evolution of all
homogeneous universes. We will also investigate the quantity $P$, because
this quantity tells us something of the physical properties of the universe
model under consideration and the general properties of the evolution of
anisotropic universes.
We will discuss, unless stated otherwise, only spatially homogeneous cosmologies
with a non-tilted $\gamma $-law perfect fluid. We also assume that the
matter sources obey the strong energy condition; the effects from
inflationary fluids are considered elsewhere \cite{wcc2}. We will investigate
all the possible spatially homogeneous models. Of special interest are the
eight Thurston geometries \cite{thurston97} which are related to the
classification of topologies in three dimensions \cite{thurston}. These
models correspond to the Bianchi types in a rather interesting way \cite
{as,fik,Kodama1}. The possibility of compactification of these Thurston
models has induced a renewed interest in the homogeneous models of Thurston
and Bianchi type \cite{BK1,BK2,Kodama2,sigBI}.

This paper is organised as follows. In section \ref{FRW} we investigate all
the models that have the FRW models as a special case. The remaining
Thurston models are investigated in section \ref{Thurston}, and in section 
\ref{NonThurston} we investigate the remaining Bianchi models of
non-Thurston type. We summarise our results in the final section.\emph{\ }

\section{Models with maximally symmetric Thurston geometries}
\label{FRW}
First, we will consider models which contain the FRW universes as
special cases. The spatial sections in these cases are either $\mb{E}^{3},\mb{H}^{3},$
or $S^{3}$. In the isotropic case these correspond to the Bianchi types I
and VII$_{0},$ V and VII$_{h}$, and IX, which contain the $k=0,-1,$ and $%
+1$ FRW universes respectively.

\subsection{The case $\mb{E}^{3}$}

For the flat case, $^{(3)}R=0$, we have already exhaustively examined the
Bianchi type I models in earlier papers \cite{wcc1,wcc2}. An important class
of solutions  is provided by the Kasner vacuum
solutions: 
\[
ds^{2}=-dt^{2}+t^{2p_{1}}dx^{2}+t^{2p_{2}}dy^{2}+t^{2p_{3}}dz^{2} 
\]
The exponents obey $\sum_{i}p_{i}=\sum_{i}p_{i}^{2}=1$. The solution space
of these solutions is a circle, and  are usually
referred to as the \emph{Kasner circle}. If\footnote{%
The metric has permutation symmetry \cite{sigDiscrete}, so any of the other
exponents could equally well be 1.} $p_{1}=1$, then the line element is
Taub's form of flat spacetime. This form of flat spacetime will be referred
to as $T$, and lies on the Kasner circle. Note that $T$ is one of two
Locally Rotationally Symmetric (LRS) spacetimes on the Kasner circle. All
vacuum Kasner solutions have maximal expansion-normalised shear: 
\[
\Sigma =1. 
\]

We will assume a perfect fluid with equation of state linking the pressure $%
p $ and density $\rho :$ 
\[
p=(\gamma -1)\rho . 
\]
For all $\gamma <2$ it can be shown that the Kasner circle is a past
attractor\footnote{%
This is because the shear invariant $\sigma ^{\mu \nu }\sigma _{\mu \nu }$,
can be viewed  as a $\gamma=2$ fluid in the Friedmann equation (\ref{friedmann}).
Hence, if the model contains a perfect fluid with $\gamma <2$, the shear
will dominate the initial singularity. This need not to be the case
when $\gamma=2$.}. Calculating the Weyl tensor, we
find that near the initial singularity 
\[
\left( C^{\alpha \beta \gamma \delta }C_{\alpha \beta \gamma \delta }\right)
_{I}\propto t^{-4} 
\]
where $t$ is the cosmological time. The quantities $P$ and $\mathcal{S}$
defined in eqs. (\ref{weyl1}) and (\ref{S}), can be shown to behave as
(assuming a Kasner-like behaviour) 
\begin{eqnarray}
P_{I} &\propto &t^{\gamma -2} \\
\mathcal{S}_{I} &\propto &t^{\gamma -1}  \nonumber  \label{eq:BtI}
\end{eqnarray}
to leading order near the initial singularity. We note that $P$ will diverge
near the initial singularity for all $\gamma <2$; $\mathcal{S}$ will
approach zero for $\gamma >1$ and diverge for $\gamma <1$. If a fluid with $%
\gamma \geq 1$ is assumed to dominate near the initial singularity then $%
\mathcal{S}$ will increase as $t\rightarrow 0$. Radiation fluids will
dominate the early history of our universe so long as particle interaction
and scattering times are shorter than the expansion time. This may not be
the case for asymptotically-free interactions between $10^{14}-10^{19}GeV$
but is likely to be restored (subject to boundary effects on the
long-wavelength part of the momentum distribution due to finite horizon
effects and the breakdown of statistical mechanics) on approach to the
Planck scale. Also, it is most likely that any self-interacting
scalar fields present will become dominated by their kinetic energies as $%
t\rightarrow 0$. This is equivalent to domination by a $\gamma =2$ fluid.
Thus an effective $\gamma >1$ in the neighbourhood of the initial
singularity, is quite probable. It might not be the case if the potential
energy of a scalar field dominated or scaled in proportion to its kinetic
energy as $t\rightarrow 0$. An example of the latter type is provided by a
scalar field $\varphi $ with an exponential potential $V(\varphi )=V_{0}\exp
[-\lambda \varphi ].$

We see that for the point $T$ on the Kasner circle, we have 
\begin{eqnarray}
\left(C^{\alpha\beta\gamma\delta}C_{\alpha\beta\gamma\delta}\right)_T &
\propto & t^{-4}\cdot t^{2(2-\gamma)}  \nonumber \\
P_T &\propto & constant  \nonumber \\
\mathcal{S}_T &\propto & t
\end{eqnarray}
near the initial singularity.

When we include a matter source with anisotropic pressure created by photons or asymptotically-free particles in an anisotropic universe
(since they are expected to be collisionless significantly below the Planck
temperature) or a cosmic magnetic field, the behaviour near the
initial singularity turns out to be quite complex. The behaviour of a
Bianchi type I model with a magnetic field \cite{LeBlanc} or
Yang-Mills field \cite{jbjl} is \emph{chaotic}.
Similar behaviour are found in Bianchi type IX and VIII. In the types VIII and IX, curvature effects cause the universe to
``bounce'' between different Kasner states; the magnetic field in the type I
case, tends to mimic this effect. The Bianchi type I with pure magnetic
field was investigated using the dynamical systems approach in \cite{LeBlanc}%
. When the universe is in a Kasner epoch, the Weyl curvature invariant
behaves as in the Kasner universe. The question to consider is: what happens
to the Weyl curvature invariants during the bounces?

As we approach the initial singularity, the duration of the bounces will
become negligible compared to the duration of the Kasner epochs in between
the bounces. We can write the Weyl curvature invariant as \cite{DynSys} 
\[
C^{\alpha \beta \gamma \delta }C_{\alpha \beta \gamma \delta }=H^{4}\mathcal{%
W}^{2} 
\]
where $\mathcal{W}$ is the dimensionless Weyl curvature invariant, and
$H=\theta/3$ is the Hubble factor. The
quantity $\mathcal{W}^{2}$ can be written so that it is only a polynomial
(of degree 4) in the state space variables. The state space for Bianchi type
I with a magnetic field is compact, and hence, all the variables are
bounded and $\mathcal{W}^{2}$ must therefore also be bounded \emph{at all times%
}. Similarly, 
\[
R^{\mu \nu }R_{\mu \nu }=H^{4}\mathcal{R}^{2},
\]
where $\mathcal{R}^{2}$ is a positive definite polynomial of the bounded
state-space variables. When the universe is in a Kasner epoch, $\mathcal{R}$
will vary as\footnote{%
When an expression like this contains two terms, we should only consider the
term which is the dominant one in each situation. Sometimes the real
expression contains cross-terms which will lie in-between the two. These
terms will be omitted in this paper, mainly for brevity.} 
\[
\mathcal{R}_{I}^{2}=\mathcal{O}(t^{2(2-\gamma )})+\mathcal{O}(t^{\frac{4}{3}%
(1-2\cos \phi )}) 
\]
where $\phi $ is an angular variable on the Kasner circle. The first mode $%
t^{2(2-\gamma )}$ comes from the perfect fluid term, while $t^{\frac{4}{3}%
(1-2\cos \phi )}$ comes from the energy density of the magnetic field. The
behaviour of $P$ during these periods will therefore vary according to the
value of the Kasner parameter $\phi $. The quantity $\mathcal{S}$ will
behave as 
\[
\mathcal{S}_{I}=\mathcal{O}(t^{\gamma -1})+\mathcal{O}(t^{\frac{1}{3}%
(1+4\cos \phi )}) 
\]
during the Kasner epochs. If $\gamma >1$ this quantity will be increasing
during this period. If $\gamma <1$ there will be Kasner epochs where $%
\mathcal{S}$ will decrease, and others where it will increase. On
``average'', assuming that all the Kasner epochs occur with equal
probability, the quantity $\mathcal{S}$ will increase. During a bounce, the
magnetic field will experience a sharp peak in its energy density. This
makes $\mathcal{R}$ experience a peak as well. Hence, during this bounce,
the quantity $\mathcal{S}$ will fall for a brief period, until it rises to a
new Kasner level again. Thus, during  Kasner oscillations, $\mathcal{S}$
will both decrease and increase in time. However, on average, $\mathcal{S}$
will increase continuously during these oscillations.\footnote{Note
that in physically realistic universes, which begin their cosmic
evolution at the Planck time, there is time for very few oscillations
to occur in the universe's history.}

The Bianchi type VII$_0$ universe model, the other Bianchi model
which is compatible with $\mb{E}^3$ spatial sections, has further interesting properties. For example, if $1\leq\gamma\leq 4/3$ the solutions of this model
will at late times isotropise in terms of the shear,
but not isotropise with respect to the Weyl
tensor \cite{WHU,NHW}. It can be shown that if $\gamma \leq 4/3$
\beq 
\Sigma_{VII_0} &\lra& 0 
\eeq 
while for $\gamma>4/3$ the shear will oscillate
as $t\lra \infty$ \cite{BS1}. The Weyl tensor has a quite complicated behaviour;
if $\gamma> 1$ then $\mc{W}$ will oscillate with larger and larger
amplitudes. The oscillations will diverge at late times in the sense
that\footnote{The authors in \cite{WHU,NHW} define a ``Weyl
parameter'' $\mc{W}$ differently from that defined here. This makes
their parameter increase monotonically at late times, while ours
will oscillate.}
\beq
\underset{t\longrightarrow\infty}{{\mathrm{lim\, sup}}}|\mc{W}^2|=\infty
\eeq
 This happens because the universe at late times
will oscillate around an isotropic state. As the universe expands, the
amplitude of the oscillations will be smaller and smaller, and hence, will
isotropise with respect to the shear. But the oscillations will become
more and more rapid while $\mc{W}$ increases beyond
bound \cite{WHU,NHW}. 

The scalar $\mathcal{W}$ measures the absolute value of Weyl curvature,
while the quantity $P$ measure the ratio of the ``matter curvature'' to the
Weyl curvature. Strictly speaking, the measure $P$ will oscillate as the
quantity $\mathcal{W}$ does,\emph{\ }but we can avoid this by either
adopting the definition of the Weyl parameter as in \cite{WHU,NHW}, or we
can take the absolute value of $\mathcal{W}^{2}$ and thereafter average over
a complete period. We will do the latter, and denote
these entities with a bar, for example $\bar{P}$. In the case of VII$_{0}$
we then get 
\[
\bar{P}_{VII_{0}}\longrightarrow \infty ,\quad \text{as }\quad
t\longrightarrow \infty 
\]
So in this case, even the ``$P$-version'' of the WCC holds. Since the
universe is ever expanding, $\mathcal{S}$ will diverge at late times as
well, as 
\[
\bar{\mathcal{S}}_{VII_{0}}\longrightarrow \infty ,\quad \text{as }\quad
t\longrightarrow \infty . 
\]

\subsection{The hyperbolic spaces $\mb{H}^{3}\ $}

We will now consider the $\mb{H}^{3}$ cases. They \ have $^{(3)}R<0.$ The future
asymptotics for these models are more easily obtained because they approach
vacuum solutions in the future. All of the type V and VII$_{h\neq 0}$
solutions will, assuming that the matter content obeys the strong energy
condition, evolve asymptotically towards plane-wave spacetimes \cite{BS}.
For plane waves, all the curvature scalars are zero even though the metrics
are anisotropic. In general, these open universe solutions will not
have a zero Weyl tensor (except for a set of zero measure which correspond
to the isotropic vacuum Milne universes). However, the
late-time behaviour of VII$_{h}$ models is to approach particular exact
solutions first found by Lukash \cite{Lukash} which are plane waves \cite{BS,BS1}.\footnote{%
There also exists an $h=4/11$ Lukash vacuum solution found by Lukash \cite
{Lukash} and a $\gamma =2$ stiff fluid solution by Barrow \cite{barrowVIIh}.}%

This behaviour indicates that the evolution of the Weyl scalar and $\mathcal{S}$
in this case is more subtle as we approach the future asymptote. The Weyl
scalars will tend to zero even though the anisotropy freezes in with $\Sigma 
$ approaching a constant value. Clearly, the Weyl scalar cannot capture the
anisotropic plane-wave modes of the gravitational field. This can be
considered as an objection to the argument that it is some function of the
Weyl scalar which is the proper measure for the entropy. We need a more
discriminating measure for the gravitational entropy, if it exists. It might
also happen that these plane-wave spacetimes are isentropic modes
corresponding to adiabatic perturbations of the gravitational field but we
would need to know whether they are extendible into hermitian modes of
a quantum cosmology.

In the V and VII$_{h}$ cases containing a perfect fluid, part of the
Kasner circle will be a past attractor for all equation of state parameters $%
\gamma <2$. Hence, the discussion for the type I model near the initial
singularity also applies to these models.

As we already have mentioned, the future attractor is a Lukash plane-wave
solution with $\Sigma \neq 0$ if $2/3<\gamma <2$ and the open FRW model with 
$\Sigma =0$ if $\gamma \leq 2/3$. Using the exact solutions in the type V
model with perfect fluid (see \cite{DynSys}), we can show that when $2/3\leq
\gamma \leq 2$ the late-time behaviour for these solutions has a Weyl
curvature invariant evolution 
\beq
\left( C^{\alpha \beta \gamma \delta }C_{\alpha \beta \gamma \delta }\right)
_{V}=t^{-4}\cdot \left[ \mathcal{O}(t^{-8})+\mathcal{O}(t^{-3\gamma -2})%
\right] .
\label{eq:WeylV}\eeq
The first of these terms comes from pure shear term effects in the absence
of matter. The second term is driven by the matter interactions. In the
presence of a perfect fluid, the pure shear term will always be sub-dominant
for $2/3\leq \gamma <2$. Under the assumption $2/3\leq \gamma \leq 2$ at
late times we have 
\begin{eqnarray}
P_{V} &\propto &t^{-\frac{3}{2}(2-\gamma )}  \nonumber \\
\mathcal{S}_{V} &\propto &t^{\frac{3}{2}\gamma }.  \label{eq:PSV}
\end{eqnarray}
Even though this is derived from the exact solution of type V, one can show
that eqs.(\ref{eq:WeylV}) and (\ref{eq:PSV}) will also hold in the generic
case (see Appendix A); the expression for Weyl curvature invariant in the
type V case is remarkable simple. Thus, in the type V case, $P$ is always
decreasing and $\mathcal{S}$ is always increasing. Note also that this
increase is due to the anisotropic modes; we assumed the universe to be
homogeneous.

The case VII$_{h}$ is more difficult. Investigations by Barrow and Sonoda 
\cite{BS1,BS}, and later by Hewitt and Wainwright \cite{HW,DynSys}
indicate that if $2/3<\gamma\leq 2 $ then the generic solution asymptotes a
plane-wave solution at late times. The plane-wave solutions have $\Sigma
=constant$ and zero Weyl curvature invariant. The approach to the plane-wave
asymptote, is subtle and a careful investigation of the Weyl curvature
invariant in a neighbourhood of the plane-wave solutions is needed. The
investigation and calculation is somewhat tedious, but the result is rather
interesting: 
\beq
\left( C^{\alpha \beta \gamma \delta }C_{\alpha \beta \gamma \delta }\right)
_{VII_{h}}\propto t^{-4}\cdot \left[ \mathcal{O}\left( t^{-\frac{2(1+\Sigma
_{+})}{1-2\Sigma _{+}}}\right) +\mathcal{O}\left( t^{-\frac{4\Sigma
_{+}+(3\gamma -2)}{1-2\Sigma _{+}}}\right) \right] .
\label{eq:WeylVII}\eeq
(see Appendix A on the nature of these asymptotes). Here, $\Sigma _{+}$ is
a constant parameter which tells us which plane-wave solution is approached 
as $t\rightarrow \infty ,$ and $\Sigma _{+}$ obeys the inequality 
\[
-\frac{1}{4}(3\gamma -2)<\Sigma _{+}<0.
\]
The shear is related to $\Sigma _{+}$ for these plane-wave solutions via the
simple relation 
\[
\Sigma _{\text{plane-wave}}=|\Sigma _{+}|.
\]
Note the appearance of the shear-to-volume expansion rate parameter in the
power of time rather than as an additive perturbation. This is a signal of
the stability of the isotopic solution being decided at second (or higher)
order because the Lukash plane-waves are exact solutions of both the
Einstein equations and the linearised Einstein equations. Thus a
linearisation of the Einstein equations about the Lukash solution gives a
zero eigenvalue \cite{BS}. Which of the two terms that in eq. (\ref
{eq:WeylVII}) is dominant depends on the parameters $\Sigma _{+}$ and $\gamma $. The
Ricci scalar squared will in general decay as 
\[
\left( R^{\mu \nu }R_{\mu \nu }\right) _{VII_{h}}\propto t^{-4}\cdot t^{-2%
\frac{4\Sigma _{+}+(3\gamma -2)}{1-2\Sigma _{+}}}.
\]
Hence, 
\[
P_{VII_{h}}=\mathcal{O}\left( t^{\frac{3(\Sigma _{+}+(\gamma -1))}{1-2\Sigma
_{+}}}\right) +\mathcal{O}\left( t^{\frac{4\Sigma _{+}+(3\gamma -2)}{%
2(1-2\Sigma _{+})}}\right),
\]
and the volume expands as $\sqrt{|g|}\propto t^{3/(1-2\Sigma _{+})}$, so $\mathcal{S}$
will increase as 
\[
\mathcal{S}_{VII_{h}}=\mathcal{O}\left( t^{\frac{3(\Sigma _{+}+\gamma )}{%
1-2\Sigma _{+}}}\right) +\mathcal{O}\left( t^{\frac{4\Sigma _{+}+3\gamma }{%
2(1-2\Sigma _{+})}}\right) .
\]
Note that both $P$ and $\mathcal{S}$ $\rightarrow \infty $ as $%
t\longrightarrow \infty $. Both the Weyl invariant and the Ricci invariants
approach zero, but the Weyl invariant does so at a slower rate than the
Ricci invariant. Ricci has $R^{\mu \nu }R_{\mu \nu }\propto \rho ^{2}$,
while the Weyl tensor contains shear modes as well. 

The work of Collins and Hawking \cite
{CH} is of relevance here. This was a study of the stability of isotropic ever-expanding
universes against spatially homogeneous perturbations of Bianchi type VII$_h$
and was not restricted to perfect fluids and comoving fluids. In the case of 
VII$_{h}$ perturbations to open FRW\ universes, isotropy was shown to be
unstable. However, this well known result requires careful interpretation.
The definition of stability used was asymptotic stability in the sense of
Lyapunov; that is, as $t\rightarrow \infty $ we need to have  $%
\Sigma \rightarrow 0.$ However, in general we have found that these VII$_{h}$
perturbations of the open FRW universe approach the Lukash plane-wave
spacetimes which have $\Sigma =$ constant ($\Sigma\rightarrow\infty$
is forbidden by $\rho>0$). Thus although isotropy is not
asymptotically stable it is stable in the sense that the deviations from
isotropy ($\Sigma =0$) are always bounded \cite{jb,bt}. Note that the
strong theorem of Collins and Hawking, assuming only the strong energy
condition ($\gamma >2/3)$ is a reflection of the property that 'the matter
doesn't matter' as $t\rightarrow \infty $ in the open universe
case. Thus, the
 behaviour determining the asymptotic stability is that of the vacuum
solution and so does not have any significant impact upon why the
present anisotropy is so 
low (see \cite{jb2} for a discussion of how this may be explained). By
contrast, in the case of type VII$_{0}$ perturbations around the flat FRW
model, the matter content of the universe is very influential and $\Sigma $
 decays to zero as $t\rightarrow \infty $ only if the
dominant form of matter is a perfect fluid with zero pressure ($\gamma =1$%
) \cite{CH}. Again, this is not very relevant to the present state of
the universe 
since in the absence of inflation the bulk of its expansion e-folds since
the Planck epoch have been in a radiation-dominated state for which isotropy
is not stable in the Collins and Hawking sense.

The space $\mb{H}^{3}$ is infinite, and in the above cases we have assumed that
we are dealing with an infinite and unbounded space
$\mb{H}^{3}$. However, if we 
relax the assumption of \emph{global} homogeneity and isotropy, we can get
other spacetimes whose spatial section at each instant of
time is a compact hyperbolic manifold (CHM). Interestingly, our earlier
discussion must be completely revised in the compact case. The compact
hyperbolic spaces are heavily constrained by Mostow's rigidity theorem; no
anisotropic deformations of CHMs are allowed. All the Bianchi models of type
V and VII$_{h\neq 0}$ with compact spatial sections have to be (locally)
isotropic \cite{BK1}. Hence, their Weyl tensors identically
vanish.\footnote{Inhomogeneous deformations are still allowed, but the
spherical symmetry of $%
\mb{H}^{3}$ is broken at the global scale when we go to its compact quotients.
Thus, the inhomogeneous Tolman-Bondi model, which we discussed in
refs. \cite{wcc1,sigLT}, 
must also be constrained.
The matter content of the Tolman-Bondi models is determined by the
dust-density $\rho (r,t)$. If $\rho $ is allowed to vary only as a function
of the radius $r$ inside a spherical shell of radius $r_{inj}$, where $%
r_{inj}$ is the coordinate \emph{injectivity radius} of the CHM, then the
compactification can be done. Outside the shell of radius $r_{inj}$, the
density $\rho $ must be constant as a function of $r$ so that the
homogeneity can be ``restored'' outside this shell. The same arguments as
for the general Tolman-Bondi model can now be made, and we get a $\mathcal{S}%
\propto t^{3}$ increase inside $r_{inj}$.} 

A more general investigation of the behaviour of $\mathcal{S}$ for CHMs is
dependent on a calculation of the inhomogeneous eigenmodes of the CHM.
This has not been done by date, but some numerical studies have been made \cite{Inoue}.

\subsection{The case $S^3$}

Turning our attention to the closed universe case, the Bianchi type IX model
has the possibility of chaotic dynamics. A closed 
universe will recollapse to a second singularity after a finite amount of
proper time has elapsed if the matter obeys the strong energy condition, the
positive pressure criteria, the dominant energy condition, and a matter
regularity condition \cite{BGT,BT}. There can be an asymmetry
between behaviour at the initial and final singularities which reflects the
relative likelihood of different initial conditions. A special initial state
that is chosen close to isotropy will deviate increasingly from isotropy on
approach to any final singularity and exhibit chaotic spacetime oscillations
induced by the anisotropy of the three-curvature. However, a
quasi-isotropic  initial
state can be made general by the assumption that the matter content at the
initial singularity is dominated by the influence of a $\gamma =2$ fluid or
a scalar field with zero potential \cite{barrowVIIh}. In this case the
difference between the quantity $\mathcal{W}$ and $P$ is also
manifest. The quantity $%
P$ is regular everywhere for the type IX case, except at the singularities.
However, at the turning-point of the volume expansion $H=0$, and so  $%
\mathcal{W}$ will diverge there.

When it comes to the initial singularity, the Bianchi type IX has a very
peculiar behaviour. Near the initial singularity it displays chaotic
behaviour in vacuum and with $\gamma \neq 2$ perfect fluids \cite
{Belinsky:1970ew,jb22,jb3,cb,dem,DH2000,dh2}. There is the issue of
the coordinate covariance of measures of chaos to be dealt with here,
but this problem can be overcome \cite{CL}. The geometrical picture is
similar to that of the magnetic Bianchi type I universe which we have already
discussed. The difference is in the Ricci tensor and in the matter terms.
The Weyl curvature invariant will have the same property as in the magnetic
Bianchi type I case. If the universe has a perfect fluid with equation of
state parameter $\gamma $ then we would expect the same behaviour as in the
Kasner universes. Hence, 
\[
\mathcal{S}_{IX}\propto t^{\gamma -1} 
\]
during the Kasner epochs. The bounces induce downward spikes in $\mathcal{S}
$, but on average we will have a steady increase of $\mathcal{S}$ with time
until dust domination or potential dominated scalar-field evolution ($\gamma
<1$ effectively).

\section{Other models of Thurston type}

\label{Thurston}

\subsection{The case $\widetilde{SL(2,\mb{R})}$}

Cosmological models with spatial sections $\widetilde{SL(2,\mb{R})}$ can be
invariant under two different Bianchi types, namely III and VIII. If we were
to assume a type III invariance with $\widetilde{SL(2,\mb{R})}$ spatial
sections, we would find that this is only compatible if the type III were
part of a larger symmetry group. Hence, type III can be considered as a
special case of VIII.

Let us therefore investigate models of Bianchi type VIII. When we compactify
the spatial space, this model becomes the most generic of the (locally)
homogeneous models. The number of free parameters can be arbitrarily large
as the topology of the space increases in complexity \cite{BK1,BK2,Kodama2}.
In addition to this, the type VIII case is interesting for its dynamical
behaviour as well. Not only does vacuum and perfect fluid ($\gamma \neq 2$) type
VIII display chaotic behaviour near the initial singularity, as in the IX
case, it can also display oscillatory behaviour at late times as in the VII$%
_{0}$ evolution \cite{yves,Ringstrom}. So, in this case, two rather peculiar
features are manifested together, and by understanding the Bianchi type VIII
could possibly be one of the keys for understanding the behaviour of $%
\mathcal{S}$ and the Weyl tensor in more generic models.

Also the perfect fluid type VIII model has a similar behaviour as the type
VII$_{0}$ \cite{wainwright}. For $1\leq \gamma <2$ the generic solution will
approach the type III form of flat space, but it does so in a way such that%
\footnote{%
Strictly speaking, these results are not rigorously proven; according
to Wainwright they rely on a combination of stability analysis and numerical
simulations \cite{wainwright}.} 
\[
\underset{t\longrightarrow \infty }{{\mathrm{lim\,sup}}}|\mathcal{W}%
^{2}|=\infty . 
\]
The shear has the asymptotic value $\Sigma =\frac{1}{4}$ in this case.
Thus for the type VIII evolution we have 
\[
P_{VIII}\approx P_{IX},\quad \mathcal{S}_{VIII}\approx \mathcal{S}_{IX}\quad 
\text{as }t\longrightarrow 0 
\]
and for $1\leq \gamma <2$ (actually for $4/5<\gamma <2$) 
\[
\bar{\mathcal{W}}_{VIII},\ \bar{P}_{VIII},\ \bar{\mathcal{S}}%
_{VIII}\longrightarrow \infty \quad \text{as }t\longrightarrow \infty . 
\]
For $2/3<\gamma \leq 4/5$, both $\mathcal{W}$ and $\Sigma $ are finite and
non-zero at late times; the solutions asymptote to the Collins type VI$_{-1}$
solution (see section \ref{sec:VIh}). We refer to refs \cite{wainwright} and \cite
{yvesPhD} for details.

\subsection{The case \textsf{Sol}}

\textsf{Sol}vegeometry has the Bianchi Lie algebra VI$_{0}$. The solutions
will have the Collins VI$_{0}$ solution as a future attractor when $%
2/3<\gamma <2$. This solution is given by 
\[
ds^{2}=-dt^{2}+t^{2}dx^{2}+t^{\frac{2-\gamma }{\gamma }}\left(
e^{2cx}dy^{2}+e^{-2cx}dz^{2}\right) 
\]
where $c\equiv\frac{\sqrt{(2-\gamma )(3\gamma -2)}}{2\gamma }$ and $\rho =\frac{%
(2-\gamma) }{\gamma ^{2}}t^{-2}$. The shear is given explicitly by 
\[
\Sigma =\frac{(3\gamma -2)^{2}}{16}. 
\]
We can also calculate the Weyl tensor for this spacetime: 
\[
\left( C^{\alpha \beta \gamma \delta }C_{\alpha \beta \gamma \delta }\right)
_{VI_{0}}=\frac{2}{3}\frac{(2-\gamma )(3\gamma -2)^{2}(4-5\gamma )}{\gamma
^{4}t^{4}}. 
\]
The Ricci squared scalar is 
\[
\left( R^{\mu \nu }R_{\mu \nu }\right) _{VI_{0}}=\frac{(2-\gamma
)^{2}(3\gamma ^{2}-6\gamma +4)}{\gamma ^{4}t^{4}} 
\]
and hence, 
\[
P_{VI_{0}}^{2}=\frac{(3\gamma -2)^{2}(4-5\gamma )}{(2-\gamma )(3\gamma
^{2}-6\gamma +4)}. 
\]
Unfortunately, $P^{2}<0$ (the magnetic part of the Weyl curvature dominates
the electric part) for $\gamma >4/5,$ which makes $P$ imaginary. Note also
that $P=0$ for $\gamma =4/5$. Ignoring the sign (taking the absolute value)
we can make $P^{2}$ non-negative for all $\gamma $. In this case, we can
define $\mathcal{S}$: 
\[
\mathcal{S}_{VI_{0}}\propto t^{\frac{2}{\gamma }}. 
\]
Hence, since this is a future attractor solution, the generic solution will
behave similarly at late times.

The generic solution starts as a special Kasner solution, and thus $P$ and $\mathcal{S}$ behave as in the Bianchi type I
case near the initial singularity.

\subsection{The case \textsf{Nil}}

\textsf{Nil}-geometry is invariant under the Bianchi type II group. The
future attractor for generic $\gamma $-law perfect fluid solutions with $%
2/3<\gamma <2$ is the Collins-Stewart solution: 
\[
ds^{2}=-dt^{2}+t^{\frac{2-\gamma }{\gamma }}\left( dx+\frac{c}{2\gamma }%
zdy\right) ^{2}+t^{\frac{2+\gamma }{2\gamma }}(dy^{2}+dz^{2}) 
\]
where $c\equiv\sqrt{(2-\gamma )(3\gamma -2)}$ and $\rho =(6-\gamma )/(4\gamma
^{2}t^{2}).$

The shear and the Weyl curvature invariant can be readily calculated 
\begin{eqnarray}
\Sigma _{II} &=&\frac{(3\gamma -2)^{2}}{64}  \nonumber \\
\left( C^{\alpha \beta \gamma \delta }C_{\alpha \beta \gamma \delta }\right)
_{II} &\propto &t^{-4}.
\end{eqnarray}
Unfortunately, the Weyl proportionality factor can be negative, positive or
zero. Taking the absolute value of the Weyl curvature invariant we obtain 
\begin{eqnarray}
P_{II} &=&constant  \nonumber \\
\mathcal{S}_{II} &\propto &t^{\frac{2}{\gamma }}.
\end{eqnarray}
The generic solution near the initial singularity will be approximated by a Kasner solution.

\subsection{The case $\mb{H}^{2}\times \mb{R}$}

\label{sec:III} This model corresponds to the Bianchi type III model. In the
perfect fluid case, the nature of the future attractor depends on the
equation of state parameter $\gamma $. If $\gamma <1$, the future attractor
is one of the Collins perfect fluid solutions. If $\gamma \geq 1$ then the
generic solutions are future asymptotic to the type III plane-wave solution.
Note also that the type III solutions can be considered as a special case of
the VI$_{h}$ solutions for $h=-1$. The VI$_{h}$ universes will in general be
of non-Thurston type and will be considered in section \ref{sec:VIh}.

When $\gamma \geq 1$, the solutions asymptote to a plane-wave solution and
so we expect that the Weyl tensor decays faster than $t^{-4}$. Not many
exact solutions are known in this case, but the LRS dust and radiation cases
are known explicitly. Actually, in the LRS case with $\gamma >1$ we do not
need the exact solutions to study the late-time behaviour. We can show that
(see Appendix A) if $1<\gamma \leq 3/2$, at late times the Weyl tensor
will decay as 
\[
\left( C^{\alpha \beta \gamma \delta }C_{\alpha \beta \gamma \delta }\right)
_{III,LRS}\propto t^{-4}\cdot t^{-4(\gamma -1)}.
\]
The late-time asymptotics still have shear and this is given by $\Sigma =1/4
$. The Ricci tensor will decay similarly, and so 
\begin{eqnarray}
P_{III,LRS} &=&constant  \nonumber \\
\mathcal{S}_{III,LRS} &\propto &t^{2}.
\end{eqnarray}
The radiation case can be derived explicitly because the exact solutions are
known.
For $3/2\leq \gamma <2$ the Weyl tensor decays as 
\[
\left( C^{\alpha \beta \gamma \delta }C_{\alpha \beta \gamma \delta }\right)
_{III,LRS}\propto t^{-4}\cdot t^{-2}, 
\]
$P$ and $\mathcal{S}$ vary as 
\begin{eqnarray}
P_{III,LRS} &\propto &t^{2\gamma -3}  \nonumber \\
\mathcal{S}_{III,LRS} &\propto &t^{2\gamma -1}.
\end{eqnarray}
Hence, both $P$ and $\mathcal{S}\rightarrow \infty $ as
$t\rightarrow\infty$ for $3/2<\gamma \leq2$.

In the LRS dust case, $\gamma =1$, the Weyl curvature invariant will
approach zero at late times as 
\[
\left( C^{\alpha \beta \gamma \delta }C_{\alpha \beta \gamma \delta }\right)
_{III,LRS}\propto t^{-4}\cdot (\ln t)^{-2}. 
\]
The Ricci tensor also decreases at the same rate, so at late times $P$ and $%
\mathcal{S}$ vary as 
\begin{eqnarray}
P_{III,LRS} &=&constant  \nonumber \\
\mathcal{S}_{III,LRS} &\propto &t^{2}\ln t.
\end{eqnarray}

The non-LRS case is slightly more troublesome. The type III dynamics can be
considered as a special case of VI$_{h}$ with $h=-1$. The generic case is
therefore considered in the VI$_{h}$ section below. The LRS type III case
is, as we explain in detail in Appendix A, a special case, and the
behaviour of the Weyl curvature invariant is  different in the
general case than for the LRS type III case. \emph{\ }If we want to
compactify the type III geometry, we have to take an LRS model. The
compactification procedure requires a higher symmetry than the
three-dimensional type III symmetry. Hence, compact spatial geometry implies
LRS.

In the case $\gamma <1$, the future attractor will be the Collins VI$_{h}$
perfect fluid solutions with $h=-1$ (see section \ref{sec:VIh}). For this
solution, the Weyl tensor decays as 
\[
\left( C^{\alpha \beta \gamma \delta }C_{\alpha \beta \gamma \delta }\right)
_{III}\propto t^{-4}
\]
while the shear is constant: 
\[
\Sigma _{III}=\frac{1}{4}(3\gamma -2)^{2}.
\]
The Ricci square decays at the same rate, so 
\begin{eqnarray}
P_{III} &=&constant,  \nonumber \\
\mathcal{S}_{III} &\propto &t^{\frac{2}{\gamma }}.
\end{eqnarray}

Near the initial singularity, the solution is approximately that of a Kasner
solution. The LRS type III will approach the Taub point $T$ on the Kasner
circle. The analysis for the LRS type III shows that near the initial
singularity, we get to leading order 
\begin{eqnarray}
\left( C^{\alpha \beta \gamma \delta }C_{\alpha \beta \gamma \delta }\right)
_{III,LRS} &\propto &t^{-4}\cdot t^{2(2-\gamma )}  \nonumber \\
P_{III,LRS} &=&constant  \nonumber \\
\mathcal{S}_{III,LRS} &\propto &t.
\end{eqnarray}
Hence, the type III behaves like the Bianchi type I universe near the
initial singularity.

\subsection{The case $S^{2}\times \mb{R}$}

This includes the closed Kantowski-Sachs case and there is no corresponding
Bianchi model. However, all the Kantowski-Sachs solutions are asymptotic to
a plane symmetric Kasner model (not $T$) as $t\rightarrow 0$. Hence, near
the initial singularity, we have 
\[
\left( C^{\alpha \beta \gamma \delta }C_{\alpha \beta \gamma \delta }\right)
_{KS}\propto t^{-4}
\]
The Ricci square is determined by the square of the energy density, and
hence, as $t\rightarrow 0$
\begin{eqnarray}
P_{KS} &\propto &t^{\gamma -2}  \nonumber \\
\mathcal{S}_{KS} &\propto &t^{\gamma -1}.
\end{eqnarray}
The Kantowski-Sachs models will recollapse for many types of fluids \cite
{BGT,BT}. Recent observations do suggest however, that we live in a state of
accelerated expansion. This type of behaviour can be explained by the
late-time influence of a quintessence scalar field. In these cases, the KS
universe can be ever-expanding. The KS models with a cosmological constant ($%
\gamma =0$) contain an unusual unstable particular solution which does not
approach the de Sitter solution as $t\rightarrow \infty $, \cite{web}. The
cosmic no hair theorem \cite{Wald} is not violated because the 3-curvature is positive.
This unusual behaviour has been used to argue for forms of dimensionally
selective inflation \cite{lz}, however, the behaviour is an artefact of the $%
S^{2}\times \mb{R}$ symmetry and is unstable even within the KS class of
solutions \cite{BYear}.

\section{Models of non-Thurston type}

\label{NonThurston}

\subsection{The Bianchi type VI$_{h}$.}

\label{sec:VIh} For this model, there exist future-attractor solutions for
all $\gamma $ (as in the VI$_{0}$ case). The nature of these attractors
depends on the group parameter $h$ (which is non-positive) and on the
equation of state parameter $\gamma $. If $\gamma $ satisfies 
\beq
-h<\frac{2-\gamma }{3\gamma -2}
\label{eq:VIh}\eeq
then the future attractor solution is the Collins VI$_{h}$ perfect fluid
solution with metric 
\[
ds^{2}=-dt^{2}+t^{2}dx^{2}+t^{\frac{2-\gamma }{\gamma }}e^{rx\frac{2-\gamma 
}{\gamma }}\left( t^{\frac{rc}{\gamma }}e^{2cx}dy^{2}+t^{-\frac{rc}{\gamma }%
}e^{-2cx}dz^{2}\right) ,
\]
where $c\equiv \sqrt{(2-\gamma )(3\gamma -2)}$ and $0<r<1$ is related to $h$
by $h=-(2-\gamma )^{2}r^{2}/c^{2}$. These solutions are universes with shear
and some of their physical properties have been studied by one of us in
connection with primordial nucleosynthesis \cite{jbmn}. The shear is found
to be 
\[
\Sigma _{VI_{h}}=\frac{1}{16}(1-3h)(3\gamma -2)^{2}.
\]
The Weyl tensor in this case has a similar behaviour as in the type VI$_{0}$
case. For some values of $\gamma $ it is negative, and others positive.
Ignoring the sign, as we did in the VI$_{0}$ case, we get 
\begin{eqnarray}
\left( C^{\alpha \beta \gamma \delta }C_{\alpha \beta \gamma \delta }\right)
_{VI_{h}} &\propto &t^{-4}  \nonumber \\
P_{VI_{h}} &=&constant  \nonumber \\
\mathcal{S}_{VI_{h}} &\propto &t^{\frac{2}{\gamma }}.
\end{eqnarray}

If the equation of state parameter $\gamma $ fails to obey eq. (\ref{eq:VIh}%
), and so 
\beq
-h\geq \frac{2-\gamma }{3\gamma -2},
\label{eq:planewavecond}\eeq
then the solutions appear to be asymptotic to plane-wave spacetimes\footnote{%
This is not rigorously proven, but Hewitt and Wainwright \cite{HW,DynSys}
have conjectured that, except for sets of measure zero, these plane-wave
solutions are global future attractors. The same is also the case for types
IV and VII$_{h}$.}. The plane-wave solutions have zero Weyl-curvature
invariants, therefore we need to know how these solutions approach the
future attractor. Unfortunately, only a few exact VI$_{h}$ solutions with
this behaviour are known.

Nevertheless, we can show that when $\gamma $ obeys the inequality (\ref
{eq:planewavecond}),  the behaviour of a generic solution is 
\[
\left( C^{\alpha \beta \gamma \delta }C_{\alpha \beta \gamma \delta }\right)
_{VI_{h}}\propto t^{-4}\cdot \left[ \mathcal{O}\left( t^{-\frac{2(1+\Sigma
_{+})}{1-2\Sigma _{+}}}\right) +\mathcal{O}\left( t^{-\frac{4\Sigma
_{+}+(3\gamma -2)}{1-2\Sigma _{+}}}\right) \right] 
\]
where $\Sigma _{+}$ characterises the different plane-wave solutions
(see Appendix A). The shear is directly related to this parameter via $\Sigma
=|\Sigma _{+}|$. The Ricci square goes as $\rho ^{2}$, and hence 
\[
P_{VI_{h}}=\mathcal{O}\left( t^{\frac{3(\Sigma _{+}+(\gamma -1))}{1-2\Sigma
_{+}}}\right) +\mathcal{O}\left( t^{\frac{4\Sigma _{+}+(3\gamma -2)}{%
2(1-2\Sigma _{+})}}\right) .
\]
Interestingly, in general $P$ will grow unbounded as we approach the future
attractor. Thus, in these solutions, the Ricci tensor will decrease faster
than the shear modes captured by the Weyl curvature invariant. The Weyl
tensor will decrease towards a plane-wave state, but as it does so, the
Ricci tensor will be sub-dominant compared to the Weyl tensor.

The volume expands as $\sqrt{|g|}\propto t^{3/(1-2\Sigma_+)}$, thus $\mathcal{S}$
will increase as 
\begin{eqnarray}
\mathcal{S}_{VI_h}=\mathcal{O}\left(t^{\frac{3(\Sigma_++\gamma)}{1-2\Sigma_+}%
}\right)+ \mathcal{O}\left(t^{\frac{4\Sigma_++3\gamma}{2(1-2\Sigma_+)}%
}\right).
\end{eqnarray}

As mentioned earlier, these solutions are future attractor solutions, and
hence, the generic solutions will have a similar behaviour for their Weyl
tensors. Near the initial singularity, the generic behaviour is Kasner-like
and the Weyl tensor behaves  as in the Bianchi type I.

\subsection{The exceptional case}

The exceptional case of Bianchi type VI$_{-1/9}^{\ast }$,  has to be
treated  separately. The exceptional case has as many free parameters
as the whole family VI$_{h}$ with arbitrary $h$ because the choice $h=-1/9$
results in the disappearance of some of the Einstein constraint equations.
Unfortunately, the exceptional case has not been much studied in the
literature. Stability properties were studied by Barrow and Sonoda \cite{BS}%
, a study of the consequences for primordial nucleosynthesis was made in 
\cite{jbmn}, and Hewitt \cite{Hewitt}, studied the evolution in a subspace
of the whole dynamical system of VI$_{-1/9}^{\ast }$ (namely the subspace
with $n_{~\alpha }^{\alpha }=0$). In this subspace Hewitt and Barrow and
Sonoda showed that for $2/3<\gamma <10/9$, $\gamma =10/9$ and $10/9<\gamma <2
$ the future attractors were the Collins VI$_{-{1}/{9}}$ perfect fluid
solutions, Wainwright $\gamma =10/9$ solutions \cite{Wainwright10-9} with
metric

\[
ds^{2}=-dt^{2}+t^{2}dx^{2}+t^{\frac 25}\left(e^{Ax}dy+wt^{\frac
45}dx\right)^{2}+t^{\frac 65}e^{-4Ax}dz^{2}
\]
where $w^{2}=9r^{2}/4-1$ and $A=r\sqrt{6}/5,$ and the Robinson-Trautman vacuum
solution\footnote{%
This solution is also sometimes called the Collinson-French solution. It
contains no free parameters. } with metric \cite{BS}

\[
ds^{2}=-dt^{2}+\frac{75}{8}t^{2}dx^{2}+\frac{5\sqrt{10}}{8}%
e^{-x}t^{\frac 65}dxdy+\frac{3}{16}e^{-2x}t^{\frac
25}dy^{2}+e^{4x}t^{\frac 65}dz^{2},
\]
respectively. We can readily calculate the Weyl tensors for these
spacetimes. For $2/3<\gamma \leq 10/9$ we get 
\[
\left( C^{\alpha \beta \gamma \delta }C_{\alpha \beta \gamma \delta }\right)
_{VI_{-1/9}^{\ast }}\propto t^{-4}.
\]
The Ricci tensor decays at the same rate, and hence 
\begin{eqnarray}
P_{VI_{-1/9}^{\ast }} &\propto &constant  \nonumber \\
\mathcal{S}_{VI_{-1/9}^{\ast }} &\propto &t^{\frac{2}{\gamma }}.
\end{eqnarray}

The Robinson-Trautman solution corresponds to a universe with shear $\Sigma
=1/3$, but has a zero Weyl curvature scalar. However, when $10/9<\gamma <2$
we can show that for any generic solution that approaches this solution, we
have 
\[
\left( C^{\alpha \beta \gamma \delta }C_{\alpha \beta \gamma \delta }\right)
_{VI_{-1/9}^{\ast }}\propto t^{-4}\cdot \left[ \mathcal{O}\left( t^{-\frac{2%
}{3}}\right) +\mathcal{O}\left( t^{-\frac{9\gamma -10}{3}}\right) \right] 
\]
(see Appendix B). 
Note that the two terms are equally dominant when $\gamma =4/3$ (radiation).
At late times the Ricci square decays faster towards zero than the Weyl curvature
invariant: 
\begin{eqnarray}
P_{VI_{-1/9}^{\ast }} &\propto &\mathcal{O}\left( t^{\frac{9\gamma -11}{3}%
}\right) +\mathcal{O}\left( t^{\frac{9\gamma -10}{6}}\right)   \nonumber \\
\mathcal{S}_{VI_{-1/9}^{\ast }} &\propto &\mathcal{O}\left( t^{3\left(
\gamma -\frac{28}{45}\right) }\right) +\mathcal{O}\left( t^{\frac{3}{2}%
\left( \gamma +\frac{4}{45}\right) }\right) .
\end{eqnarray}
In this case the Weyl curvature invariant will dominate at late times. At
early times, the generic solution approaches a certain point on the Kasner
circle (not $T$), and hence, the analysis of Bianchi type I applies again.

As mentioned, this analysis is just for a special subspace in the dynamical
state space of VI$_{-1/9}^{\ast }$. Nobody, at least to the authors'
knowledge, has made a complete analysis of the whole dynamical behaviour of
the exceptional case. However, page 171 in \cite{DynSys} appears to indicate
that the exceptional case has a chaotic initial singularity. The future
attractor seems to be a self-similar spacetime, and hence the above results
may hold in that case. Further investigation of this case is required. 

\subsection{The Bianchi type IV}

It is conjectured that generically the Bianchi type IV solutions move
asymptotically towards plane-wave solutions. The behaviour is similar to the
VI$_{h}$ case, and the late-time behaviour for the Weyl tensor is similar to
the plane-wave case. Hence, we will have 
\[
\left( C^{\alpha \beta \gamma \delta }C_{\alpha \beta \gamma \delta }\right)
_{IV}\propto t^{-4}\cdot \left[ \mathcal{O}\left( t^{-\frac{2(1+\Sigma _{+})%
}{1-2\Sigma _{+}}}\right) +\mathcal{O}\left( t^{-\frac{4\Sigma _{+}+(3\gamma
-2)}{1-2\Sigma _{+}}}\right) \right] 
\]
(see Appendix A). Again the shear is given by $\Sigma =|\Sigma _{+}|$. The
Ricci square goes as $\rho ^{2}$, and hence 
\[
P_{VI_{h}}=\mathcal{O}\left( t^{\frac{3(\Sigma _{+}+(\gamma -1))}{1-2\Sigma
_{+}}}\right) +\mathcal{O}\left( t^{\frac{4\Sigma _{+}+(3\gamma -2)}{%
2(1-2\Sigma _{+})}}\right) .
\]
Finally, we have 
\[
\mathcal{S}_{VI_{h}}=\mathcal{O}\left( t^{\frac{3(\Sigma _{+}+\gamma )}{%
1-2\Sigma _{+}}}\right) +\mathcal{O}\left( t^{\frac{4\Sigma _{+}+3\gamma }{%
2(1-2\Sigma _{+})}}\right) .
\]
So, in this case the $P$ and $\mathcal{S}$ $\rightarrow \infty $ at late
times. Near the initial singularity, the type IV behaves like a Kasner
universe, and hence, the Bianchi type I analysis applies again.

\section{Conclusion}

In this paper we have investigated the general behaviour of the Weyl
curvature invariant for spatially homogeneous universes containing a perfect
fluid as $t\rightarrow 0$ and $t\rightarrow \infty $.  For all the models,
except for the chaotic ones, the solutions asymptote in the past to points
on the Kasner circle. The Kasner universe, which is a special vacuum case of
the Bianchi type I model, has been treated exhaustively elsewhere \cite
{sigBI,wcc1}.

In the chaotic models, we saw that the Weyl curvature invariant
$C^{\alpha \beta \gamma \delta }C_{\alpha \beta \gamma \delta }$, the
invariant $P^2= \frac{C^{\alpha \beta \gamma \delta }C_{\alpha \beta
\gamma \delta }}{R^{\mu\nu}R_{\mu\nu}}$, and the invariant $%
\mathcal{S}=\sqrt{|g|}P$ experience peaks and sinks. On the average, however, $\mathcal{S%
}$  increases steadily as the universe expands from the initial singularity.

The late-time behaviour was found to fall into one of five categories:

\begin{enumerate}
\item  \textbf{Ricci dominance.} This category includes models for which $P$
is decreasing at late times, indicating that the Ricci square dominates over
the Weyl curvature invariant asymptotically.

\item  \textbf{Weyl-Ricci balanced.} In this category $P$ is approximately
constant at late times. The Ricci and Weyl tensors are approximately
proportional at late times.

\item  \textbf{Weyl dominance.} In this category $P\rightarrow \infty $ as $%
t\rightarrow \infty $, but the expansion-normalized Weyl tensor, $\mathcal{W}^2=C^{\alpha \beta \gamma \delta }C_{\alpha \beta \gamma \delta }/H^4$, will
decrease. Except in the exceptional VI$_{-1/9}^*$ case, these models evolve
towards vacuum plane-wave spacetimes, but the Weyl curvature invariant
decreases more slowly than the Ricci square.

\item  \textbf{Extreme Weyl dominance.} In these models, $\mathcal{W}$ $%
\rightarrow \infty $ as $t\rightarrow \infty $.

\item  \textbf{Recollapsing models.} These models have no late-time
behaviour, they recollapse after a finite time subject to the usual
technical energy conditions required for recollapse \cite{BGT}.
\end{enumerate}

\begin{table}[tbp]
\centering
\begin{tabular}{|c|c|c|}
\hline
Type & Matter & Category \\ \hline\hline
I & $2/3<\gamma \leq 2$ & 1 \\ 
& $2/3<\gamma \leq 4/3 +\pi_{\mu\nu}$ & 1 \\ 
& $4/3<\gamma \leq 2 + $ mag.field & 2 \\ \hline
II & $2/3<\gamma<2 $ & 2 \\ \hline
III$_{LRS}$ & $2/3<\gamma\leq 3/2 $ & 2 \\ 
& $3/2 <\gamma\leq 2 $ & 3 \\ \hline
IV & $2/3 <\gamma\leq 2$ & 3 \\ \hline
V & $2/3 <\gamma\leq 2$ & 1 \\ \hline
VI$_0$ & $2/3 <\gamma< 2$ & 2 \\ \hline
VI$_h$ & $2/3 <\gamma<\frac{2(1-h)}{1-3h} $ & 2 \\ 
& $\frac{2(1-h)}{1-3h}\leq\gamma \leq 2$ & 3 \\ \hline
VI$^*_{-1/9}$ & $2/3 <\gamma \leq 10/9$ & 2 \\ 
& $10/9<\gamma \leq 2$ & 3 \\ \hline
VII$_0$ & $2/3 <\gamma < 1 $ & 1 \\ 
& $1<\gamma\leq 2$ & 4 \\ \hline
VII$_h$ & $2/3 <\gamma\leq 2 $ & 3 \\ \hline
VIII & $2/3 <\gamma <4/5$ & 2 \\ 
& $4/5 <\gamma \leq 2 $ & 4 \\ \hline
IX & $2/3 <\gamma \leq 2 $ & 5 \\ \hline
KS & $2/3 <\gamma \leq 2 $ & 5 \\ \hline
I-IX & $0\leq \gamma <2/3$ & 1 \\ \hline
\end{tabular}
\caption{Classification of the late-time behaviour. The matter content
of the universe is labelled by the perfect fluid equation of state
parameter, $\gamma$; $\pi_{\mu\nu}$ indicates the presence of
trace-free anisotropic stresses and type I also includes the case of a
pure magnetic field.}
\label{results}
\end{table}
In the Table \ref{results} we have summarised our results, and in the
rightmost column we have indicated to which of the five categories each of
the spatially homogeneous models belong. We note that category-3 cosmologies
are not often discussed in the literature\footnote{%
This behaviour seems to have been noted for a special class of the VI$_{h}$
solutions in \cite{GCW}.}. All of the plane-wave solutions of class B belong
to this category. Note also that when we compactify the topology of the
spatial sections \cite{BK1,BK2}, most of the models in this
category disappear in the compactification procedure; types IV and VI$_{h}$
cannot be compactified, type III must be locally rotationally
symmetric (LRS), and VII$_{h}$ must be
isotropic. However, some of the type III spacetimes (namely those LRS with $%
\gamma >3/2$) in category 3 can be compactified.

A general criterion to decide when a dynamical system will end up in category
3 can be given. If the late-time asymptote is a vacuum spacetime which is
not conformally flat, but nevertheless has a zero Weyl curvature invariant,
then the quantity $P$ will generically increase without bound. If, in
addition, the state space of the corresponding dynamical system is also
compact, then $\mathcal{W}$ has to be bounded (and thus decreasing), and
hence, the dynamical system will end up in category 3. The reason for this
is as follows. For a conformally flat spacetime, both the Weyl curvature
invariant and the Weyl tensor itself vanish. Hence, any linear perturbation
of this conformally flat solution will produce linear corrections to the
components of the Weyl tensor. Since the linear terms are the lowest-order
terms in the Weyl tensor, quadratic terms will be the lowest-order terms in
the Weyl curvature invariant. For non-conformally flat spacetimes, the Weyl
curvature tensor will not be zero. Hence, the linear terms will not be the
lowest-order terms in the perturbation. Thus, in the Weyl curvature
invariant some of the linear terms will survive even though the lowest-order (constant) terms could vanish identically (which happens for the
plane-wave solutions). The electric and the magnetic parts of the Weyl
tensor are not zero for plane-wave solutions, they just happen to be of
equal magnitude and cancel when the invariant is formed.

If the late-time asymptote is conformally flat, but not Ricci flat (hence,
non-vacuum), then the system will typically end up in category 1. There is
also a very small chance that the spacetime will end up in category 4 if the
state space is non-compact. If the asymptote is both Ricci flat and
conformally flat, all five categories are possible. If the state space is
compact, then category 4 is ruled out. The fifth category depends to a
certain extent on the geometry of the late-time attractor. For instance, 
inflationary spacetimes with $\rho +3p<0$, are all governed by the ``no
hair'' theorem which states that the late-time asymptote is a conformally
flat non-vacuum spacetime. Hence, all ever-inflating cosmologies should end
up in category 1.

In all of the models investigated, except for those in category 5, $\mathcal{%
S}$ increases at late times. Hence, $\mathcal{S}$ behaves as expected by the
WCC. It is not clear whether we should always expect a Weyl 'entropy' to increase
if the initial conditions are close to some state of effective gravitational
equilibrium. Of course, our study has only discussed spatially homogeneous
universes and it remains to be seen whether small density and
gravitational wave inhomogeneities evolve in a manner consistent with a
thermodynamic interpretation of the Weyl curvature and whether the inclusion of
significant inhomogeneities, black holes, and damping processes introduces
significant new ingredients to the search for the elusive gravitational
'entropy'.  

\section*{Appendix A: The Weyl-Curvature Invariant for Bianchi types of
class B.}

We will briefly show how the Weyl curvature invariant behaves near the
plane-wave equilibrium points of type III, V, VI$_{h}$, IV and VII$_{h}$.
The discussion is based on Hewitt and Wainwright's approach \cite{DynSys,HW}
(see also \cite{BS} and \cite{BogN,bog}).

For all the plane-wave solutions we will have $C_{\alpha \beta \gamma \delta
}C^{\alpha \beta \gamma \delta }=0$, so the question we would like to answer
is how the Weyl curvature invariant approaches zero at late times. The Weyl
curvature invariant is given explicitly by 
\[
C_{\alpha \beta \gamma \delta }C^{\alpha \beta \gamma \delta }=8H^{4}\left[ (%
\mathcal{E}_{AB}\mathcal{E}^{AB}+6\mathcal{E}_{+}^{2})-(\mathcal{H}_{AB}%
\mathcal{H}^{AB}+6\mathcal{H}_{+}^{2})\right] 
\]
where the expansion-normalised electric parts and magnetic parts of the Weyl
curvature are given by 
\begin{eqnarray}
\mathcal{E}_{+} &=&\Sigma _{+}(1+\Sigma _{+})-\tilde{\Sigma}+\frac{2}{3}%
N_{+}^{2}-\frac{2}{3}\tilde{h}\tilde{A}  \nonumber \\
\mathcal{H}_{+} &=&3\Delta   \nonumber \\
\mathcal{E}_{AB}\mathcal{E}^{AB} &=&6\bigg[\tilde{\Sigma}(2\Sigma
_{+}-1)^{2}+4\left\{ \tilde{A}\Sigma _{+}+\frac{1}{3}(N_{+}^{2}-\tilde{h}%
\tilde{A})(\tilde{A}+(N_{+}^{2}-\tilde{h}\tilde{A})-2\tilde{\Sigma})\right\} 
\nonumber \\
&&+4\Delta ^{2}-4N_{+}\Delta (2\Sigma _{+}-1)+\frac{4}{3}\tilde{h}\tilde{A}%
(N_{+}^{2}-\tilde{h}\tilde{A})\bigg]  \nonumber \\
\mathcal{H}_{AB}\mathcal{H}^{AB} &=&6\big[3\Sigma _{+}^{2}(N_{+}^{2}-\tilde{h%
}\tilde{A})+2\tilde{\Sigma}(N_{+}^{2}-\tilde{h}\tilde{A})+\tilde{\Sigma}%
\tilde{A}+6\Delta ^{2}  \nonumber \\
&&+12\Sigma _{+}N_{+}\Delta +4\tilde{h}\tilde{A}\tilde{\Sigma}\big].
\end{eqnarray}
The shear is given by 
\[
\Sigma =\widetilde{\Sigma }+\Sigma _{+}^{2}.
\]
All the variables and their interpretations are explained in detail in refs. 
\cite{DynSys,HW}. In particular, $\tilde{h}$ is related to the Lie algebra
parameter $h$ via $\tilde{h}=1/h$. The Bianchi types are related to $\tilde{h%
}$ as follows. 
\begin{eqnarray}
\tilde{h}<0: &&\text{VI}_{h}  \nonumber \\
\tilde{h}=0: &&\text{IV}  \nonumber \\
\tilde{h}>0: &&\text{VII}_{h}. 
\end{eqnarray}
Note that type III is VI$_{-1}$ and type V is contained in the discussion as
the special case of $\tilde{h}=0$.

In these variables the plane-wave equilibrium points are given by 
\begin{eqnarray}
\tilde{\Sigma} &=&-\Sigma _{+}(1+\Sigma _{+}),\quad \Delta =0,\quad \tilde{A}%
=(1+\Sigma _{+})^{2},  \nonumber \\
N_{+}^{2} &=&(1+\Sigma _{+})[\tilde{h}(1+\Sigma _{+})-3\Sigma _{+}],
\label{planewave}
\end{eqnarray}
where $\Sigma _{+}$ is a parameter which satisfies 
\begin{eqnarray}
&-1<\Sigma _{+}\leq \frac{\tilde{h}}{3-\tilde{h}},&\text{ \ if \ }\tilde{h}<0 
\nonumber \\
&-1<\Sigma _{+}<0,&\text{\ if \ }\tilde{h}\geq 0
\end{eqnarray}
The eigenvalues of the dynamical system for these equilibrium points are
given by 
\beq
\{0,\quad -4\Sigma _{+}-(3\gamma -2),\quad -2[(1+\Sigma _{+})\pm iN_{+}]\},
\label{eigenval}\eeq
where $N_{+}$ is given by (\ref{planewave}). The equilibrium points are
future attractors whenever 
\[
\Sigma _{+}>-\frac{1}{4}(3\gamma -2).
\]

\begin{figure}[tbp]
\centering
\epsfig{file=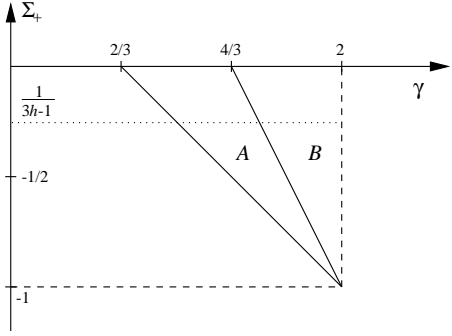, width=6cm}
\caption{In region $A$ the eigenvalue $-4\Sigma_+-(3\protect\gamma-2)$ is
larger than $-2(\Sigma_++1)$. In region $B$, $-2(\Sigma_++1)$ is the
largest. }
\end{figure}

The second of these eigenvalues is directly related  to the time evolution
of the perfect fluid. If $\Omega $ is the expansion-normalised energy
density, $\Omega $ will in a close neighbourhood of the plane-wave
equilibrium points obey 
\[
\Omega ^{\prime }=[-4\Sigma _{+}-(3\gamma -2)]\Omega
\]
where prime denotes derivation with respect to the time variable
$\tau$ defined by 
\[
\frac{dt}{d\tau }=\frac{1}{H}.
\]
For the plane-wave solutions, we have $\Omega =0$. The dependence of the
eigenvalues on the equation of state parameter is shown in Figure 1.

Upon a small variation of the variables from these equilibrium points, the
variation will decay as linear combination of factors $\exp (\lambda
_{i}\tau )$ where $\lambda _{i}$ are the eigenvalues (\ref{eigenval}).
Hence, we would naively expect the Weyl curvature to behave as 
\[
\mathcal{W}^{2}=\mathcal{O}(\exp (\lambda _{i}\tau )).
\]
But here we have to be careful. It can be shown that the generic behaviour
of the Weyl curvature invariant is like this, but in some important cases it
is not. For instance, for the radiative solution of LRS type III which we
mentioned in section \ref{sec:III}, all the linear terms from the variation
around the plane-wave solution vanish. If we constrain ourself to lie
in the
invariant subspace $\Delta =\Sigma _{+}N_{+},\ 3\Sigma _{+}^{2}+\tilde{h}%
\tilde{\Sigma}=0,\ \tilde{A}>0$, then we would get 
\begin{eqnarray}
\mathcal{W}^{2} &=&\frac{9}{(3-\tilde{h})}\bigg[(1+\tilde{h})\left\{ (12%
\tilde{h}+7)\Omega +(69\tilde{h}-121)\sigma _{+}\right\} +\sqrt{1+\tilde{h}}%
(27\tilde{h}+5)\nu   \nonumber \\
&&+(\text{quadratic terms in }\Omega ,\ \sigma _{+},\ \nu )\bigg].
\end{eqnarray}
Here, $\Omega $ is a small variation of the density of the perfect fluid,
and $\sigma _{+}$ and $\nu $ are small variations of $\Sigma _{+}$ and $N_{+}
$ respectively. The LRS type III is precisely this invariant subspace when $%
\tilde{h}=-1$. Note that if and only if $h=\tilde{h}=-1$, the linear terms
in $\Omega $, $\sigma _{+}$ and $\nu $ will vanish identically. In the case
were $\gamma <3/2$, $\Omega $ will dominate at late times over $\sigma _{+}$
and $\nu $, thus 
\[
\mathcal{W}^{2}\propto \Omega ^{2}
\]
at late times. Note also that $R^{\mu \nu }R_{\mu \nu }\propto H^{4}\cdot
\Omega ^{2}$ which means that $P_{III}\approx constant$ at late times for
these LRS type III solutions.

The type V model is of special interest. The invariant subspace is given by $%
\tilde{h}=0$, $\tilde{A}>0$ while $\Sigma _{+}=\Delta =N_{+}=0$. In this
case the Weyl curvature invariant is particularly simple: 
\[
\mathcal{W}_{V}^{2}=48(2\tilde{\Sigma}^{2}+\tilde{\Sigma}\Omega ).
\]

For completeness, we should mention that for all of the plane-wave solutions
we have a simple inversion time evolution for the mean Hubble parameter. The
simple Milne evolution arises only when the shear parameter $\Sigma _{+}=0:$
\begin{eqnarray}
t &\propto &e^{(1-2\Sigma _{+})\tau }  \nonumber \\
H &=&\frac{1}{1-2\Sigma _{+}}t^{-1}.
\end{eqnarray}

\section*{Appendix B: The Weyl-Curvature Invariant in the exceptional case
of Bianchi type VI$_{-\frac{1}{9}}^{\ast }$}

For the dynamical system of Hewitt \cite{Hewitt}, the Weyl tensor is 
\begin{eqnarray}
\mathcal{W}^{2} &=&\frac{8}{3}\Sigma _{+}^{2}\left( 3+18\Sigma
_{12}^{2}+4\Sigma _{+}^{2}-16N_{\times }^{2}\right)   \nonumber \\
&&+32N^{2}\left( \Sigma _{+}+N_{\times }^{2}-3\Sigma _{12}^{2}\right)
+18\Sigma _{12}^{2}\left( 1+3\Sigma _{12}^{2}\right) 
\end{eqnarray}
while the shear is 
\[
\Sigma =\frac{4}{3}\Sigma _{+}^{2}+3\Sigma _{12}^{2}
\]
The Robinson -Trautman vacuum solution has 
\[
\Sigma _{+}=-\frac{1}{3},\quad \Sigma _{12}=\frac{\sqrt{5}}{9},\quad
N_{\times }=\frac{\sqrt{2}}{2}.
\]
The eigenvalues of the linear system with respect to this equilibrium point
are  given by \cite{Hewitt}, \cite{BS} 
\[
-\frac{2}{3}\left( 1\pm \sqrt{14}i\right) ,\quad \frac{10-9\gamma }{3}.
\]
In this case, the Weyl tensor will have linear terms in all of the three
variables. Thus the Weyl tensor will decay as $\exp (\lambda _{max}\tau )$
where $\lambda _{max}$ is the eigenvalue with the largest real part. The
time variable $\tau $ is related with the cosmological time via 
\[
t\propto e^{\frac{5}{3}\tau }
\]
near the equilibrium point.

\section*{Acknowledgments}

We deeply appreciate the useful and insightful comments made by {\O }. Gr\o %
n, S. Siklos and Y. Gaspar. SH was funded by the Research Council of Norway.

\end{document}